\documentstyle[twoside,fleqn,espcrc2,epsf]{article}

\newcommand{\half}
             {\mbox{\small $\frac{1}{2}$}}          
\newcommand{\smhalf}
             {\mbox{\tiny  $\frac{1}{2}$}}          
\newcommand{\quarter}
             {\mbox{\small $\frac{1}{4}$}}          
\newcommand{\third}
             {\mbox{\small $\frac{1}{3}$}}          
\newcommand{\twothird}
             {\mbox{\small $\frac{2}{3}$}}          

\newcommand{\AmS}{{\protect\the\textfont2
  A\kern-.1667em\lower.5ex\hbox{M}\kern-.125emS}}

\title{
       \vspace{-3.65cm}                                     
       {\normalsize DESY 96--127}    \\[-0.2cm]             
       {\normalsize HUB--EP--96/32}  \\[-0.2cm]             
       {\normalsize June 1996}       \\                     
       \vspace{2.25cm}                                      
       A Preliminary Lattice Study of the Glue in the Nucleon%
            \thanks{Talk presented by R. Horsley at Lat96,  
                    St. Louis, U.S.A.}}                     

\author{M.~G\"ockeler%
           \address{H{\"o}chstleistungsrechenzentrum HLRZ,
                    c/o Forschungszentrum J{\"u}lich, D-52425 J{\"u}lich,
                                                             Germany}
           \hspace{-0.2cm}
           \address{Institut f\"ur Theoretische Physik, RWTH Aachen,
                    D-52056 Aachen, Germany}
           \hspace{-0.2cm}
           \address{Institut f\"ur Theoretische Physik, J.~W. Goethe
                    Universit\"at, D-60054 Frankfurt, Germany},
        R.~Horsley%
           \address{Institut f\"ur Physik, Humboldt-Universit\"at zu Berlin,
                    Invalidenstra{\ss}e 110, D-10115 Berlin, Germany},
        E.-M.~Ilgenfritz$^{\rm d}$,
        H.~Oelrich$^{\rm a}$
           \hspace{-0.2cm}
           \address{DESY-IfH Zeuthen, D-15735 Zeuthen, Germany},
        H.~Perlt%
           \address{Fak. f. Physik und Geowiss., Universit\"at Leipzig,
                    Augustusplatz 10--11, D-04109 Leipzig, Germany},
        P.~E.~L. Rakow$^{\rm a}$ \hspace{-0.2cm} $^{\rm e}$,
        G.~Schierholz$^{\rm a}$ \hspace{-0.2cm} $^{\rm e}$
           \hspace{-0.2cm}
           \address{Deutsches Elektronen-Synchrotron DESY,
                    Notkestra{\ss}e 85, D-22603 Hamburg, Germany},
        A.~Schiller$^{\rm f}$
        and
        P.~Stephenson$^{\rm e}$}
       
\begin{document}

\begin{abstract}
About half the mass of a hadron is given from gluonic contributions.
In this talk we calculate the chromo-electric and chromo-magnetic components
of the nucleon mass. These computations are
numerically difficult due to gluon field ultra-violet fluctuations.
Nevertheless a high statistics feasibility run using quenched Wilson fermions
seems to show reasonable signals.
\end{abstract}

\maketitle

\setcounter{footnote}{0}

\section{INTRODUCTION}
\label{intro}

One of the earliest experimental indications that the nucleon consists
not only of three quarks, but also has a gluonic contribution came
from the measurement of the fraction of the nucleon momentum
carried by the quarks. That this did not sum up to $1$ as is required
from the energy--momentum sum rule gave evidence for the
existence of the gluon. Denoting $\langle x \rangle^{(f)}$ as the
fraction of the nucleon momentum carried by parton $f$ we have
\begin{equation}
   \sum_q \langle x \rangle^{(q)} + \langle x \rangle^{(g)} = 1.
\label{intro.a}
\end{equation}
Experimentally $\langle x \rangle^{(u+d)} \approx 0.4$ so the
missing component is large, at least $50\%$ of the total nucleon momentum.

We have previously estimated the quark contribution from a lattice
calculation (at least for the valence part in the quenched approximation)
\cite{goeckeler95a,goeckeler94a,goeckeler96a}. In this contribution
we shall consider%
\footnote{Some results, with smaller statistics were given in
\cite{goeckeler96a}.}
$\langle x \rangle^{(g)}$. Analogously to 
$\langle x \rangle^{(q)}$ we have, with ${\cal M}$ denoting Minkowski space
and averaging over the polarisations,
\begin{eqnarray}
   \lefteqn{ \langle N, \vec{p}
            | O^{{\cal M}(g)\mu_1\mu_2} -
              \quarter \eta^{\mu_1\mu_2} O^{{\cal M}(g)\alpha}_{
                                           \phantom{{\cal M}(g)\alpha}\alpha}
            | N, \vec{p} \rangle}
                                       \nonumber \\
    &=& 2 \langle x \rangle^{(g)} \left[ p^{\mu_1}p^{\mu_2} -
                                       \quarter \eta^{\mu_1\mu_2} m_N^2
                                  \right],
\label{intro.b}
\end{eqnarray}
with
$O^{{\cal M}(g)\mu_1\mu_2} = 
    - \mbox{Tr} F^{{\cal M}\mu_1\alpha}
                    F^{{\cal M}\mu_2}_{\phantom{{\cal M}\mu_2}\alpha}$.
(Higher moments can also be considered, by inserting covariant 
derivatives between the $F$'s, \cite{goeckeler96a}.)

\section{THE LATTICE METHOD}
\label{lattice}

We now turn to the lattice. The gluon operators are euclideanised%
\footnote{$E^{{\cal M}i} = F^{{\cal M}i0} \to iF_{i4} \equiv iE_i$ and
$B^{{\cal M}i} = -{\smhalf}\epsilon^{ijk}F^{\cal M}_{\phantom{\cal M}jk} \to
{\smhalf}\epsilon_{ijk}F_{jk} \equiv B_i$.}
and discretised in the usual way. For the field strength we choose
the usual clover leaf form%
\footnote{Note that we, like most workers in the field, do not subtract
the trace of the clover term, to make $F^{lat}$ traceless in the
colour fields. This is an $O(a^5)$ operator and so does not contribute to
the continuum identification of the clover term with $F$.},
\cite{mandula83a}, which belongs to an irreducible
representation of the $4$-dimensional hypergroup $H(4)$. Defining
$O_{\mu\nu}=-\mbox{Tr}F^{lat}_{\mu\alpha} F^{lat}_{\nu\alpha}$ this then gives
with the two obvious operator choices
\begin{eqnarray}
   O_a &=& O_{i4} = \mbox{Tr}( \vec{E} \times \vec{B} )_i,
                                       \nonumber  \\
   O_b &=& O_{44} - {\third} O_{jj}
                  = {\twothird} \mbox{Tr}( -\vec{E}^2 + \vec{B}^2 ),
\label{lattice.a}
\end{eqnarray}
($O^{\cal M}_a \to i O_a$ and $O^{\cal M}_b \to O_b$).
Both choices have their problems: Operator (a) needs a non-zero momentum
$p_i$, while operator (b) requires a delicate subtraction between two
terms similar in magnitude. The relation to $\langle x \rangle^{(g)}$
is given by
\begin{eqnarray}
   \langle N, \vec{p} | O_{Ra} | N, \vec{p} \rangle 
         &=& - 2 i E_N p_i \langle x \rangle^{(g)}_a,
                                       \nonumber  \\
   \langle N, \vec{0} | O_{Rb} | N, \vec{0} \rangle 
         &=& 2 m_N^2 \langle x \rangle^{(g)}_b,
\label{lattice.b}
\end{eqnarray}
with $O_R$ denoting the renormalised operator.

We shall not dwell here on details of the numerical calculations,
\cite{goeckeler95a,goeckeler94a}, but just mention that the method we use
to extract matrix elements from the lattice is standard,
namely evaluating the ratio of $3$-point
to $2$-point nucleon correlation functions. The $3$-point function is
easy to calculate: we simply multiply the $2$-point function with
the appropriate gluon operator for every configuration. This sits 
at time $\tau$ between the baryon--anti-baryon (placed at $t$ and $0$
respectively). $\tau$ and $t-\tau$ are (hopefully) large enough, $\ge d_m$
say, so that all the excited nucleon states have died out.
We found it expedient
to sum over all allowed values of $\tau$.
Thus we consider
\begin{eqnarray}
   R(t) &=& {\langle B(t) {1\over t-2d_m+1} \sum_{\tau=d_m}^{t-d_m}
                    O(\tau) \overline{B}(0) \rangle
            \over \langle B(t) \overline{B}(0) \rangle },
                                   \nonumber \\
        &=& {1 \over 2E_N} \langle N, \vec{p}| O | N, \vec{p} \rangle.
\label{lattice.c}
\end{eqnarray}
We work with quenched Wilson fermions at $\beta=6.0$ and
$\kappa = 0.1515, 0.1530, 0.1550$ on a $16^3\times 32$ lattice with 
antiperiodic time boundary conditions for the fermion.
We have generated O(5000) sources (on $3000$--$3500$ configurations
with Jacobi smeared source/sink).

\section{THE RAW DATA}
\label{raw}

\begin{figure}[hbt]
\vspace*{-1.25cm}
\hspace*{-1.00cm}
\epsfxsize=8.15cm \epsfbox{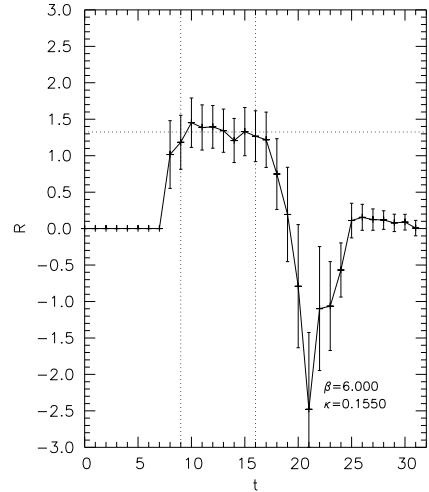}  
\vspace*{-4.00cm}
\caption{\footnotesize $R(t)$ for the vacuum subtracted operator
         $-\mbox{Tr} \vec{E}^2$. The vertical dotted lines denote the
         fit interval, the fit value being given by the horizontal
         dotted line.}
\vspace*{-0.75cm}
\label{re2}
\end{figure}

\begin{figure}[hbt]
\vspace*{-1.50cm}
\hspace*{-1.00cm}
\epsfxsize=8.15cm \epsfbox{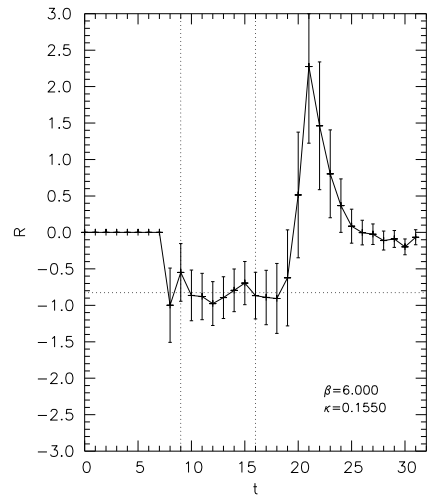}  
\vspace*{-4.00cm}
\caption{\footnotesize $R(t)$ for the vacuum subtracted operator
$\mbox{Tr} \vec{B}^2$.}
\vspace*{-0.75cm}
\label{rb2}
\end{figure}

\begin{figure}[hbt]
\vspace*{-1.50cm}
\hspace*{-1.00cm}
\epsfxsize=8.15cm \epsfbox{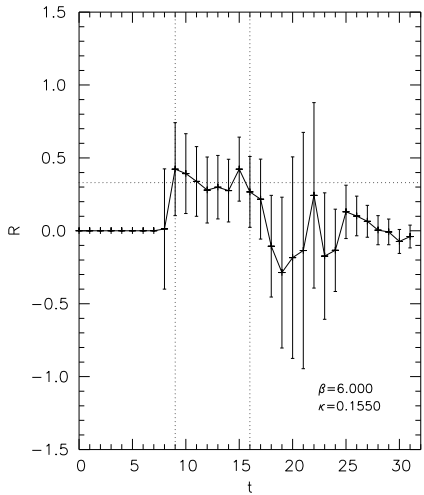}  
\vspace*{-4.00cm}
\caption{\footnotesize $R(t)$ for $O_b$.}
\vspace*{-0.75cm}
\label{rx1bg}
\end{figure}

We shall first consider $O_{44} = -\mbox{Tr} \vec{E}^2$. In Fig.~\ref{re2}
we show $R(t)$ for this operator. We fixed $d_m=4$ and hope to see a
signal after $7$ (before this there is not enough time to insert
the operator) and about $17$ (after this the nucleon mixes
with its parity partner). Indeed a signal is seen. A similar picture holds for
${\half}(O_{44}-O_{jj}) = \mbox{Tr} \vec{B}^2$ as shown
in Fig.~\ref{rb2}.
Considering $O_b$ directly is given in Fig.~\ref{rx1bg}.
As expected there is a large cancellation between the chromo-electric
and -magnetic pieces. While the error bars are uncomfortably
large, there does seem to be a signal. This is worse using $O_a$;
indeed the best we can say is that it is not inconsistent with $O_b$.

\section{RENORMALISATION}
\label{renormalisation}

As gluon operators are singlets, they can mix with the quark singlet.
But there exists a combination of singlet operators with vanishing
anomalous dimension. (This is due to the conservation of the energy-momentum
tensor.) We may estimate the renormalisation factor $Z_b$:
\begin{itemize}
\item Using first order perturbation theory. We find \cite{perlt96a}
      $Z_b = 1 + a_1 g^2 + \ldots, a_1 = 0.220$.
      We shall use this in Fig.~\ref{x1bg}.
\item Non-perturbatively. In \cite{martinelli85a} a proposal was made
      to determine $Z$ from MC simulations. We have looked at
      $\langle AOA \rangle$ for about $100$ gauge fixed configurations.
      Huge noise is present but $Z_b$ is consistent with $1$.
\end{itemize}

\section{RESULTS AND DISCUSSION}
\label{results}

\begin{figure}[hbt]
\vspace*{-1.50cm}
\hspace*{-1.00cm}
\epsfxsize=8.15cm \epsfbox{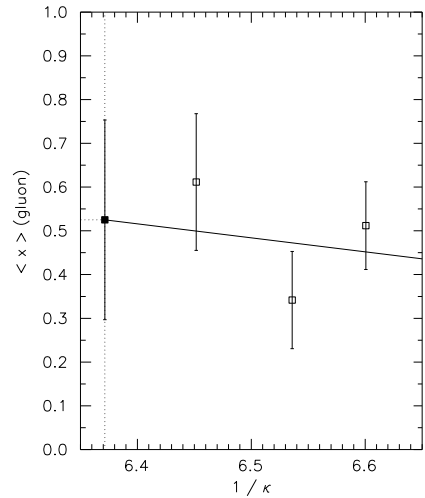}  
\vspace*{-4.00cm}
\caption{\footnotesize Chiral extrapolation for $\langle x \rangle^{(g)}_b$.}
\vspace*{-0.75cm}
\label{x1bg}
\end{figure}

Extracting $\langle x \rangle^{(g)}_b$ from $R$, we may now attempt a chiral
extrapolation. This is shown in Fig.~\ref{x1bg}. 
While the quality of the fit is not so good, the result is at least
encouraging $\approx 0.53\pm 0.23$. This is at least not inconsistent
with the expectations previously discussed. Of course our ultimate
aim is to attempt a mass splitting of the nucleon, in the same spirit as
\cite{ji95a}. This seems a difficult goal with our present method:
probably a two order of magnitude improvement
in statistics is required.

\section*{ACKNOWLEDGMENTS}
\label{acknowledgements}

The numerical calculations were performed on the APE
(Quadrics QH2) at DESY-Zeuthen
with some of the earlier computations on the Bielefeld University APE.
We thank both institutions.

\end{document}